\documentclass[conference]{IEEEtran}
\IEEEoverridecommandlockouts
\usepackage{cite}
\usepackage{amsmath,amssymb,amsfonts}
\usepackage{algorithmic}
\usepackage{graphicx}
\usepackage{textcomp}
\usepackage{xcolor}
\usepackage{comment}

\begin{document}

\title{Distributed Threat Intelligence at the Edge Devices: A Large Language Model-Driven Approach}

\author{\IEEEauthorblockN{Syed Mhamudul Hasan$^{1,2,3}$, Alaa M. Alotaibi$^{1}$, Sajedul Talukder$^{1,3}$, Abdur R. Shahid$^{1,2,3}$}
\IEEEauthorblockA{\textit{$^{1}$School of Computing, Southern Illinois University, Carbondale, IL, USA} \\
\textit{$^{2}$Secure and Trustworthy Intelligent Systems (SHIELD) Lab}\\
\textit{$^{3}$Center for Research and Education in AI and Cybersecurity (CARE-AI-C)}\\
syedmhamudul.hasan@siu.edu, alaa@siu.edu, sajedul.talukder@siu.edu, shahid@cs.siu.edu} \vspace{-1.02cm}
}

\maketitle

\begin{abstract}

With the proliferation of edge devices, there is a significant increase in attack surface on these devices. The decentralized deployment of threat intelligence on edge devices, coupled with adaptive machine learning techniques such as the in-context learning feature of Large Language Models (LLMs), represents a promising paradigm for enhancing cybersecurity on resource-constrained edge devices. This approach involves the deployment of lightweight machine learning models directly onto edge devices to analyze local data streams, such as network traffic and system logs, in real-time. Additionally, distributing computational tasks to an edge server reduces latency and improves responsiveness while also enhancing privacy by processing sensitive data locally. LLM servers can enable these edge servers to autonomously adapt to evolving threats and attack patterns, continuously updating their models to improve detection accuracy and reduce false positives. Furthermore, collaborative learning mechanisms facilitate peer-to-peer secure and trustworthy knowledge sharing among edge devices, enhancing the collective intelligence of the network and enabling dynamic threat mitigation measures such as device quarantine in response to detected anomalies. The scalability and flexibility of this approach make it well-suited for diverse and evolving network environments, as edge devices only send suspicious information such as network traffic and system log changes, offering a resilient and efficient solution to combat emerging cyber threats at the network edge. Thus, our proposed framework can improve edge computing security by providing better security in cyber threat detection and mitigation by isolating the edge devices from the network.

\end{abstract}


\section{Introduction}

Edge computing is ubiquitous in terms of privacy and security for processing data closer to the source. To augment this approach, the Large Language Model (LLM) can enhance security by identifying and mitigating potential threats. The LLM is a group of AI models that are best at certain tasks, like understanding and creating natural language and domain-specific situations, such as for personalized assistants~\cite{Wu_2023} and threat detection~\cite{motlagh2024large}. OpenAI's Generative Pre-trained Transformer (GPT), a specific LLM implementation based on the transformer architecture, has revolutionized this approach with billions of parameters. Furthermore, the in-context learning feature in GPT can leverage LLM's understanding capabilities to learn new skills within relevant and authentic linguistic contexts without full retraining or fine-tuning.

Threat intelligence at the edge refers to the practice of gathering, analyzing, and applying threat intelligence data at the periphery of a network or system, where interactions with external entities occur~\cite{conti2018cyber}. We propose a noble approach to edge threat intelligence at the edge that involves deploying lightweight AI models directly onto edge devices, such as routers, firewalls, and different IoT devices. These models can continuously observe network traffic, system logs, and configurations to identify patterns indicative of potential security threats by distributing the intelligence to different edge devices.

\begin{figure}[t]
    \centering
    \includegraphics[width = 0.47\textwidth]{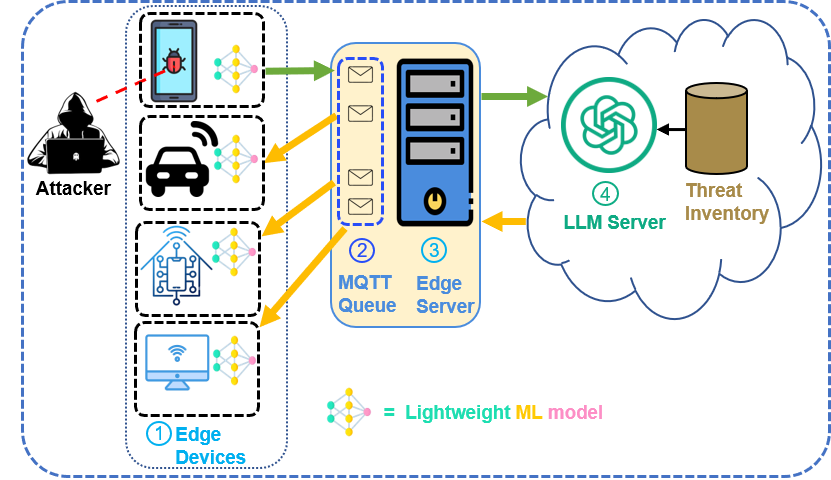}
    \vspace{-10pt}
    \caption{Configuration of the lightweight ML model at Edge devices with Edge server and Central LLM with trained with a large inventory of threat intelligence}
    
    \label{fig:system-architecture}
\vspace{-20pt}
\end{figure}

\section{System Architecture}

The key components of this approach to threat intelligence at the edge include a lightweight ML model trained to detect anomalies by analyzing network traffic and system logs. The lightweight ML model detects malicious activities as well as security threats in real-time, and communicates with the edge server by Message Queuing Telemetry Transport (MQTT) protocol. MQTT is a lightweight publish-subscribe protocol which transfers the data to the edge server and also among other edge devices and local edge, and a central LLM server trained with updated threat repositories. The central server can also temporarily update its knowledge through the in-context learning (ICL) feature of GPT. The local edge server enables edge devices, and the ML model provides edge devices with limited processing power and memory without compromising performance or resource efficiency. In figure~\ref{fig:system-architecture}, we describe the entire process by dividing into four main components:

\begin{enumerate}

\item \textbf{Edge devices with lightweight ML model:} These ML models are designed to operate efficiently on edge devices with limited computational resources, enabling real-time threat analysis and response without relying heavily on cloud-based resources. By processing threat intelligence data locally on edge devices, it can provide real-time detection of security threats analyzing network data, device logs and other parameters without introducing significant delays or latency. This allows for immediate response actions, such as blocking malicious traffic or alerting the edge server, to mitigate potential risks. This model can vary from device to device, as every device has its own thread landscape in its deployment environment.

\item \textbf{MQTT which integrates edge devices with edge servers:} Edge devices will be connected to the edge server via the MQTT channel. Also, the edge devices can communicate with each other to share data with the help of this channel, thus creating one-to-many communication.

\item \textbf{Edge server deployed locally:} The edge server provides the MQTT queue, where edge devices basically exchange data with each other and the local edge server, preserving user privacy and security without transmitting sensitive information to external cloud servers. The edge server monitors the activity and, more specifically, the warning given by the compromised edge device by alerting the system administrator and blocking the edge device from communicating with others. This decentralized approach minimizes the risk of unauthorized modifications or data breaches in two way verification. Additionally, the communication latency is minimal as, in most scenarios, the edge server and edge devices are geographically located close together, which is critical for such communication.

\item \textbf{LLM server:} The central threat intelligence solution will be trained with popular threat inventory. It can also adapt to and learn from new threats and attack vectors over time by ICL feature of LLM. Through continuous training and updates, LLM can improve its accuracy and effectiveness in identifying emerging security threats at the edge. In cases of unknown threats, the edge server, connected through a high-speed network, can communicate with the central LLM server about the type of attack, possible vulnerability, and effective solution for mitigating the risk in the shortest possible time.

\end{enumerate}
 
This approach to threat intelligence at the edge can offer a proactive and efficient means of enhancing security in distributed computing environments, where traditional centralized threat detection mechanisms may be less effective. By leveraging lightweight AI models directly on edge devices, we can detect compromised edge devices with evolving cyber threats.

\section{Methodology}

To protect the other edge devices from compromised devices in near real-time data involves multiple steps. Firstly, we assume the edge devices have network connectivity, and those are connected to the edge server via the MQTT protocol. All communication are encrypted with Secure Socket Layer (SSL) to prevent additional man-in-the-middle attacks. After receiving any messages from edge devices via the MQTT queue, the edge server analyzes the issue using a trained ML model. In an unknown case, edge server informs the central LLM server at a certain interval. The lightweight AI model implemented with TensorFlow Lite~\cite{DBLP:journals/corr/abs-2010-08678} can detect and identify any suspicious activity. After detection with other meta-information like location, severity, etc., it will alert the edge server. If the edge server finds that request, analyze it and respond to other edge devices, notifying the monitoring team. In such an environment, other edge devices will not continue to communicate with the infected devices protecting the valuable data from the attacker.

For experimentation, we chose two Raspberry Pis and one android mobile phone. For edge intelligence, we deploy a trained tensorflow model to the Raspberry Pi and convert the tensorflow ML model to the tensorflow Lite deployed on the android phone or any wide range of IoT chips and microcontrollers. As every device at the edge gets a ML model, it becomes an intelligent device. Also, we use an edge server with an Intel (R) CoreTM i7 and 64 GB of RAM, which has an MQTT server running. All the edge devices are connected to the MQTT edge server via a client application. Furthermore, the edge server is connected to a central LLM server, which we consider OpenAI's GPT API which we will train with popular threat libraries. Moreover, we propose to enhance central server intelligence through in-context learning of GPT where edge servers can update the server model through the central LLM via a REST API call.

\section{Conclusion and Future Work}

The innovative side of this approach lies in the integration of LLM and diverse other popular technologies like ML application, edge server intelligence and MQTT to tackle cybersecurity challenges specific to lightweight edge devices. The proposed framework will have a relative impact on the field of edge device security by addressing wider white box attacks like zero days attack, poisoning attack etc. considering resource limitations while maintaining data privacy and scalability. In the future, we will provide a practical demonstration of the system to demonstrate the framework's applicability in real-world deployment.

\bibliographystyle{IEEEtran}
\bibliography{main}

\end{document}